\documentclass[aps,prd,12point,twocolumn,nofootinbib,showpacs,superscriptaddress]{revtex4-1}
\def\theequation{\arabic{section}.\arabic{equation}}
\usepackage{psfrag}
\usepackage{subfigure}
\usepackage{color}
\usepackage{mathrsfs}
\usepackage{graphicx}
\usepackage{amssymb, bm}
\usepackage{amsmath, amsthm}
\usepackage{epstopdf}
\usepackage{hyperref}
\usepackage{enumerate}
\usepackage{longtable}

\newcommand{\be}{\begin{equation}}
\newcommand{\ee}{\end{equation}}

\begin{document}
\def\theequation{\arabic{section}.\arabic{equation}} 

\title{Symmetry of Brans-Dicke gravity as a novel solution-generating 
technique}

\author{Valerio Faraoni}
\email[]{vfaraoni@ubishops.ca}
\affiliation{Department of Physics and Astronomy and {\em STAR} Research 
Cluster, Bishop's University, 2600 College Street, Sherbrooke, Qu\'ebec, 
Canada J1M~1Z7 }

\author{Dilek K. \c{C}iftci}
\email[]{dkazici@ubishops.ca, dkazici@nku.edu.tr}
\affiliation{Department of Physics, Nam{\i}k Kemal University, 
Tekirda\u{g}, Turkey}
\affiliation{Department of Physics and Astronomy and {\em STAR} Research 
Cluster, Bishop's University, 2600 College Street, Sherbrooke, Qu\'ebec, 
Canada J1M~1Z7 }

\author{Shawn D. Belknap-Keet}
\email[]{sbelknapkeet02@ubishops.ca}
\affiliation{Department of Physics and Astronomy, Bishop's 
University, 2600 College Street, Sherbrooke, Qu\'ebec, 
Canada J1M~1Z7
}



\begin{abstract}

A symmetry of Brans-Dicke gravity in (electro)vacuo or in the presence of 
conformally invariant matter is presented and used as a 
solution-generating technique starting from a known solution as a seed. 
This novel technique is applied to generate, as examples, new spatially 
homogeneous and isotropic cosmologies, a 3-parameter family of spherical 
time-dependent spacetimes conformal to a Campanelli-Lousto geometry, and 
a family of cylindrically symmetric geometries.

\end{abstract}


\maketitle

\section{Introduction} 
\label{sec:1}
\setcounter{equation}{0}

There is plenty of motivation for studying theories of 
gravity alternative to General Relativity (GR), both theoretically and 
experimentally. Attempts to quantize GR 
invariably introduce modifications to it in the form of extra 
dynamical fields or higher order field equations, and these corrections 
are not necessarily Planck-scale suppressed. The prototype of the 
alternative to GR is scalar-tensor gravity. Its  
simplest incarnation is Brans-Dicke theory \cite{BD}, which was  
generalized to richer forms of scalar-tensor gravity \cite{ST}. In the 
1980s, waning 
interest in this class of theories by the gravity community was renewed by 
the realization that the simplest string theory, bosonic string theory, 
reduces 
to $\omega=-1$ Brans-Dicke gravity in the low-energy limit 
\cite{bosonic}. 

More urgent motivation comes from cosmology. The 1998 discovery 
that the expansion of the universe is accelerated can be explained  by the 
standard $\Lambda$ cold dark matter cosmological model based on GR only 
at the price of 
introducing a completely {\em ad hoc} dark energy accounting for 
approximately 70\% of the energy content of the universe 
\cite{AmendolaTsujikawabook}. A possible way to avoid introducing dark 
energy is by modifying gravity. Many theories of modified gravity have 
been studied and intense experimental and theoretical efforts aiming at 
testing gravity are underway or under planning (see the reviews 
\cite{Bertietc}). Probably the most popular class of modified gravity 
theories motivated by cosmology is $f(R)$ gravity (\cite{CCT}, see 
\cite{reviews} for reviews). $f(R)$ gravity turns out to be a Brans-Dicke 
theory in disguise, corresponding to the special value $\omega=0$ of the 
Brans-Dicke coupling and to a special potential for the 
scalar degree of freedom \cite{reviews}. Apart 
from $f(R)$ gravity, Brans-Dicke theory is the toy model of choice 
to explore deviations from GR involving scalar degrees of freedom in many 
areas, including cosmology, black holes, gravitational waves, no-hair 
theorems and ways to evade them, stealth fields, and apparent horizons. 
Older research which led to the introduction of the original Brans-Dicke 
theory involves Mach's principle \cite{BD} and Dirac's idea that the 
constants of nature may actually be dynamical fields \cite{Dirac}, which 
is partially realized in the feature of Brans-Dicke gravity that the 
effective gravitational coupling strength becomes, roughly speaking, the 
inverse of the Brans-Dicke scalar field $\phi$ \cite{BD}. There has been 
renewed interest in varying ``constants'' of physics in recent years (see 
\cite{Barrowbook} for a popular exposition). Extra motivation related to 
the quantization of gravity is provided by the finding that generalized 
Brans-Dicke solutions describe asymptotically Lifshitz black holes  
\cite{MaedaGiribet}.

Analytical solutions of scalar-tensor gravity can provide insight into 
aspects of these directions of research, but they are not as numerous as 
the better known solutions of GR \cite{KSMcCH}. It is valuable, therefore, 
to find general solution-generating techniques in scalar-tensor gravity. 
Here we focus on a symmetry group of Brans-Dicke gravity (enriched by the 
possibility of an arbitrary potential $V(\phi)$ for the Brans-Dicke scalar 
field) 
in the presence of conformally invariant matter \cite{myBDlimit}, which is 
really a 
restricted conformal invariance of the theory and is reminiscent of the 
broader conformal invariance of string theories \cite{Polchinski}. We 
explore the use of this 
symmetry as a novel technique to generate new solutions of 
Brans-Dicke gravity using known solutions as seeds. As examples of 
application of this technique, we find three different kinds of 
analytical solutions: in the cosmological context, then spherically 
symmetric and time-dependent solutions, and finally cylindrically 
symmetric geometries. We use units in which the speed of light {\em in 
vacuo} and Newton's constant are unity and we follow the notation of 
Ref.~\cite{Waldbook}.

\section{A symmetry of Brans-Dicke theory} 
\label{sec:2} 
\setcounter{equation}{0}

In this section we generalize the symmetry of Brans-Dicke theory 
with $V(\phi)=0$ found in 
\cite{myBDlimit} to the case in which the Brans-Dicke scalar field 
$\phi$ is 
endowed with a potential and conformally invariant matter is present. 
For ease of exposition, we begin with the vacuum 
theory and, in the last subsection, we include conformally invariant 
matter.

\subsection{Vacuum Brans-Dicke theory with any potential}

The action is 
\be
S_{BD}= \int d^4 x \sqrt{-g} \left[ \phi R -\frac{\omega}{\phi} \, g^{ab} 
\nabla_a \phi \nabla_b \phi -V(\phi) \right] \,.\label{BDaction}
\ee
This action is invariant in form under the operation $\left( g_{ab}, \phi 
\right) \rightarrow \left( \tilde{g}_{ab}, \tilde{\phi} \right)$, where
\begin{eqnarray}
\tilde{g}_{ab} &=& \Omega^2 g_{ab} =\phi^{2\alpha} g_{ab} \,, 
\label{symmetry1}\\
&&\nonumber\\
\tilde{\phi} &=& \phi^{1-2\alpha} \,, \label{symmetry2}
\end{eqnarray}
for $\alpha \neq 1/2$, that is, a conformal transformation of the metric 
with conformal factor $\Omega=\phi^{\alpha}$ and a non-linear redefinition 
of the scalar field. Since it is $\phi>0$ to guarantee the positivity of 
the gravitational coupling, the conformal transformation is well defined 
(except at spacetime points where $\phi$ diverges, which are to be regarded 
as physical singularities). A tilde denotes geometric quantities 
constructed with 
the conformally rescaled metric $\tilde{g}_{ab}$.  By using the well-known 
transformation properties \cite{Synge, Waldbook, Carroll, mybook} 
\begin{eqnarray}
\tilde{g}^{ab} &=& \Omega^{-2} g^{ab} \,,\\
&&\nonumber\\
\sqrt{-\tilde{g}} & = & \Omega^4  \sqrt{-g} \,,\\
&&\nonumber\\
\tilde{R} &=& \Omega^{-2} \left( R-\frac{6\Box\Omega}{\Omega}\right) \,,
\end{eqnarray}
and Eq.~(\ref{symmetry2}), one obtains
\begin{eqnarray}
R &=& \phi^{2\alpha} \tilde{R}  
- \frac{6\alpha ( 1-\alpha)}{(1-2\alpha)^2} \,  
 \phi^{6\alpha-2} \tilde{g}^{ab} 
\tilde{\nabla}_a \tilde{\phi} \tilde{\nabla}_b \tilde{\phi} \nonumber\\
&&\nonumber\\
&\, & +\frac{6\alpha}{1-2\alpha} \, 
\phi^{4\alpha-1}\tilde{\Box}\tilde{\phi} \,. \label{mmminch}
\end{eqnarray}
The term proportional to $\tilde{\Box}\tilde{\phi}$ which appears in the 
action (as  a contribution coming from $\sqrt{-g}\, \phi R$) because of 
the 
last term in the right-hand side of 
Eq.~(\ref{mmminch}) can be written as
\be
\frac{6\alpha}{1-2\alpha} \,\sqrt{-\tilde{g}} \,  
\tilde{\Box}\tilde{\phi}= 
\frac{6\alpha}{1-2\alpha} \, \partial_{\mu} \left( \sqrt{-\tilde{g}} 
\,\tilde{g}^{\mu\nu} \partial_{\nu} \tilde{\phi} \right) \,,
\ee
which is integrated to produce a boundary term giving zero 
contribution when the action is varied. This term is ignored in the 
following. The Brans-Dicke action~(\ref{BDaction}), therefore, becomes
\begin{widetext}
\be
S_{BD}= \int d^4x \sqrt{ -\tilde{g}} \left\{ \tilde{\phi} \tilde{R} 
-\left[ \frac{\omega}{(1-2\alpha)^2}+ 
\frac{6\alpha(1-\alpha)}{(1-2\alpha)^2} \right] \frac{ 
\tilde{g}^{ab}}{\tilde{\phi}} \tilde{\nabla}_a \tilde{\phi} 
\tilde{\nabla}_b \tilde{\phi} -
\tilde{\phi}^{ \frac{-4\alpha}{1-2\alpha} }V( \phi ) \right\} \,.
\ee
\end{widetext}
By redefining the Brans-Dicke coupling and scalar field potential as
\begin{eqnarray}
\tilde{\omega}( \omega, \alpha)&=& \frac{ \omega 
+6\alpha(1-\alpha)}{(1-2\alpha)^2} \,, \label{newomega}\\ 
&&\nonumber\\
\tilde{V}( \tilde{\phi}) &=& \tilde{\phi}^{\frac{-4\alpha}{1-2\alpha}} 
V\left( \tilde{\phi}^{\frac{1}{1-2\alpha}} \right) \,, \label{newV}
\end{eqnarray}
the Brans-Dicke action is rewritten as \cite{myBDlimit} 
\be
S_{BD}= \int d^4x \sqrt{ -\tilde{g}} \left[ \tilde{\phi} \tilde{R} 
- \frac{ \tilde{\omega}}{\tilde{\phi}}  
\tilde{g}^{ab}  \tilde{\nabla}_a \tilde{\phi} 
\tilde{\nabla}_b \tilde{\phi} -
\tilde{V}( \tilde{\phi} ) \right] \,,
\ee
{\em i.e.}, it is invariant in form under the 
transformation~(\ref{symmetry1}), (\ref{symmetry2}), provided that the 
changes~(\ref{newomega}), (\ref{newV}) are 
made. In addition, the transformations of the type~(\ref{symmetry1}), 
(\ref{symmetry2}) form a 1-parameter Abelian group \cite{myBDlimit}. 

As a special case, we note that a power-law potential 
\be
V(\phi)=V_0 \phi^n 
\ee
(where $V_0$ and $n$ are constants) is invariant in form, {\em i.e.}, the 
symmetry produces another power-law potential
\be
\tilde{V} ( \tilde{\phi} ) = V_0 \tilde{\phi}^{ \tilde{n}}
\ee
with the new power 
\be
\tilde{n}= \frac{n-4\alpha}{1-2\alpha} \,.
\ee
An even more special case is $n=2$, for which also the  power in the 
potential is left invariant, $\tilde{V} (\tilde{\phi}) 
= m^2 \tilde{\phi}^2/2 =V(\phi)$ and 
$\tilde{n}=n=2$ when  $V(\phi)=m^2\phi^2/2$.

\subsection{Electrovacuum Brans-Dicke theory}
\label{subsec:Maxwell}

When an electromagnetic field is present as a form of matter, the action 
is 
\begin{eqnarray}
S_{BD} &=& \int d^4 x \sqrt{-g} \left[ \phi R -\frac{\omega}{\phi} \, 
g^{ab} \nabla_a \phi \nabla_b \phi -V(\phi) \right.\nonumber\\
&&\nonumber\\
&\, & \left. -  \,  F^{ab}F_{ab}   \right] \,,
\end{eqnarray}
where $ F^{ab}$ is the Maxwell tensor. Since the latter has conformal 
weight $s=0$ \cite{Waldbook}, $\tilde{F}_{ab}=F_{ab}$ 
and 
\be
\sqrt{-g} \, F^{ab}F_{ab} = \sqrt{-\tilde{g}} \, \widetilde{F^{ab}F_{ab}} 
\,,
\ee
so that also   $ \sqrt{-g} \, {\cal L}_{(m)}$  for this form of matter 
remains invariant under the transformation~(\ref{symmetry1}), 
(\ref{symmetry2}).

\subsection{Conformally invariant matter}

It is tempting to ask whether Brans-Dicke theory is left invariant by the 
transformation~(\ref{symmetry1}), (\ref{symmetry2}) in the presence of any 
other form of matter, for example conformally invariant matter. This 
property would be especially important for applications, {\em e.g.}, in 
cosmology or in stars when a radiation  
fluid is present. The action principle for fluids is notoriously 
nontrivial \cite{Seliger, Schutz, Brown}, therefore in this case it is 
more convenient to analyze directly the transformation of the field 
equations. The variation of the Brans-Dicke action~(\ref{BDaction}) 
with the addition of a matter action produces the field equations
\begin{widetext}
\begin{eqnarray}
 R_{ab}-\frac{1}{2}\, g_{ab} R &=& 
\frac{8\pi}{\phi } \, T_{ab} + \frac{\omega }{\phi^2} \left( 
\nabla_a \phi \nabla_b \phi -\frac{1}{2}\, g_{ab} g^{cd}\nabla_c 
\phi \nabla_d \phi  \right) 
 +\frac{1}{\phi}  \left( \nabla_a \nabla_b \phi  
-g_{ab} \Box \phi \right) -
\frac{V}{2\phi} \, g_{ab} \,, \label{BDfe1}\\
&&\nonumber \\
\Box \phi &=& \frac{1}{2\omega+3}\left[ 
\frac{8\pi T}{\phi}   + \phi \, \frac{d V}{d\phi} 
-2V \right] \,, \label{BDfe2}
\end{eqnarray}
\end{widetext}
where $T_{ab}$ is the matter stress-energy tensor and $T$ is its trace. 
A rather long but straightforward calculation gives the transformation 
properties of Eqs.~(\ref{BDfe1}) and (\ref{BDfe2}) under 
the operation~(\ref{symmetry1}), (\ref{symmetry2}). The scalar field 
equation~(\ref{BDfe2}) becomes 
\be
\tilde{\Box}\tilde{\phi} = \frac{1}{2\tilde{\omega}+3}\left[ 
\frac{8\pi}{1-2\alpha} \, \tilde{\phi}^{\frac{-4\alpha}{1-2\alpha}} 
T  +\tilde{\phi} \, \frac{d\tilde{V}}{d\tilde{\phi}} -2\tilde{V} 
\right] \,,\label{BDfe2transformed}
\ee
where $\tilde{\omega}$ and $\tilde{V}( \tilde{\phi})$ are given by 
Eqs.~(\ref{newomega}) and~(\ref{newV}). Therefore, Eq.~(\ref{BDfe2}) is 
invariant in form under the 
transformation~(\ref{symmetry1}), (\ref{symmetry2}) only for conformally 
invariant matter with $T=0$.

Under the same transformation, the other field equation~(\ref{BDfe1}) 
becomes
\begin{eqnarray}
&& \tilde{R}_{ab}-\frac{1}{2}\, \tilde{g}_{ab} \tilde{R} = 
\frac{8\pi}{\tilde{\phi}^{ \frac{1}{1-2\alpha} } } T_{ab} \nonumber\\
&&\nonumber\\
&&+ \frac{ \tilde{\omega} 
}{\tilde{\phi}^2} \left( \tilde{\nabla}_a \tilde{\phi} \tilde{\nabla}_b 
\tilde{\phi} -\frac{1}{2}\, \tilde{g}_{ab} \tilde{g}^{cd} \tilde{\nabla}_c 
\tilde{\phi} \tilde{\nabla}_d \tilde{\phi}  \right) \nonumber\\
&&\nonumber\\
&&+\frac{1}{\tilde{\phi}}  \left( \tilde{\nabla}_a \tilde{\nabla}_b 
\tilde{\phi}  -\tilde{g}_{ab} \tilde{\Box}\tilde{\phi} \right) -
\frac{\tilde{V}}{2\tilde{\phi}} \, \tilde{g}_{ab} 
\,.\label{BDfe1transformed}
\end{eqnarray}
The stress-energy tensor $T_{ab}$ of matter, which by now we know is 
required to be 
conformally invariant if the operation~(\ref{symmetry1}), 
(\ref{symmetry2}) 
is imposed to be a symmetry of the theory, transforms according to 
$\tilde{T}_{ab}=\Omega^{-2} \, T_{ab}$ \cite{FujiiMaeda, Waldbook}. Then 
the first term in the right-hand side of Eq.~(\ref{BDfe1transformed}) 
becomes  $ 8\pi \tilde{T}_{ab}/\tilde{\phi}$ and the 
form of this equation 
is the same of Eq.~(\ref{BDfe1}) before the transformation. We conclude 
that Eqs.~(\ref{symmetry1}) and~(\ref{symmetry2}) describe   
a symmetry of Brans-Dicke theory in the presence of an arbitrary 
(regular) scalar field potential and of conformally invariant matter. 
Examples 
include the Maxwell field in four spacetime dimensions already mentioned 
in Sec.~\ref{subsec:Maxwell} and a radiation fluid with equation of state 
$P=\rho/3$. 

The Brans-Dicke field $\phi$ couples to the trace of the energy-momentum 
tensor of ordinary matter (cf. Eq.~(\ref{BDfe2})) and only conformally 
invariant matter is covariantly conserved after 
a conformal transformation $g_{ab} \rightarrow \tilde{g}_{ab}=\Omega^2 
g_{ab}$. In fact, as is well known in scalar-tensor gravity, the covariant 
conservation equation $\nabla^b T_{ab}=0$ becomes \cite{BD, FujiiMaeda, 
mybook}
\be
\tilde{\nabla}^b \tilde{T}_{ab}= -\tilde{T} \nabla_a \ln \Omega \,,
\ee
and only $T=0$ (which occurs if and only if $\tilde{T}=0$) guarantees 
covariant conservation after the conformal rescaling.

\section{Application to Brans-Dicke cosmology}
\label{sec:3}
\setcounter{equation}{0}

We now apply the new solution-generating technique to spatially 
homogeneous and isotropic Brans-Dicke cosmology (see \cite{mybook, 
FujiiMaeda} for reviews). In general, this symmetry is not a Noether 
symmetry \cite{booksalv} nor a Hojman symmetry \cite{CapozzielloRoshan}.   
There are indications that the symmetry does not survive Wheeler-DeWitt 
quantization in minisuperspace (at least in the spatially flat case) 
because quantum effects cause an anomalous symmetry breaking similar to 
that occurring in condensed matter systems \cite{Pal}. This fact is, 
however, immaterial in the present work, which is confined to classical 
gravity.  First we use power-law, and then exponential solutions as seeds. 
In both cases the line element is the FLRW one in comoving coordinates
\be
ds^2=-dt^2 +S^2(t) \left( \frac{dr^2}{1-kr^2} +r^2 d\Omega_{(2)}^2 \right) 
\,,
\ee
where the curvature index $k$ is normalized to $0, \pm 1$ and  $ 
d\Omega_{(2)}^2 =d\theta^2 +\sin^2 \theta \, d\varphi^2$ is the 
line element on the unit 2-sphere.

 Before proceeding we note that, in the case $\omega=-1$ 
corresponding to the bosonic string theory \cite{bosonic}, the well-known 
duality of pre-big-bang cosmology \cite{prebigbang}
\be
S \rightarrow \bar{S}=1/S \,, \;\;\;\;\;\; \phi \rightarrow \bar{\phi}=S^6 
\phi  \,,
\ee
is not reproduced by, and is unrelated to, the symmetry 
(\ref{symmetry1}), (\ref{symmetry2}) that we study in our work.

\subsection{Power-law solutions} 

We first consider vacuum 
Brans-Dicke theory with $V \equiv 0$ and we look for power-law solutions 
of the form
\begin{eqnarray}
S(t) & = & S_0 t^p \,,\\
&&\nonumber\\
\phi(t) & = & \phi_0 t^q \,,
\end{eqnarray}
where $S_0>0, \phi_0>0, p$, and $q$ are constants. Most of the known exact 
solutions of Brans-Dicke cosmology are of this form \cite{mybook}, which 
includes the Brans-Dicke dust solution \cite{BD}, the O'Hanlon and Tupper 
family \cite{OHanlonTupper}, and the Nariai family \cite{Nariai}. Here we 
consider vacuum solutions.

After the conformal transformation~(\ref{symmetry1}) with parameter 
$\alpha$, the line element reads
\be
d\tilde{s}^2 = 
-t^{2\alpha q} dt^2 +S_0^2 t^{2(p+\alpha q)} \left( \frac{dr^2}{1-kr^2} 
+r^2 d\Omega_{(2)}^2 \right) \,,\label{nonso}
\ee
where an irrelevant 
multiplicative constant has 
been dropped. We now introduce the new time 
coordinate $\tau$ defined by $d\tau = t^{\alpha q} dt$ for $q \neq 0$, or 
\be
t=\left( \alpha q+1 \right)^{\frac{1}{\alpha q+1}} \tau^{ 
\frac{1}{\alpha q+1}} \,,
\ee
with the choice of a common origin for $t$ and $\tau$ and $\alpha \neq 
-1/q, 1/2$. The line element~(\ref{nonso}) is then  written using this  
comoving time as
\be
d\tilde{s}^2 = - d\tau^2 +S_0^2 \tau^{\frac{2(p+\alpha q)}{\alpha q+1} }  
\left( \frac{dr^2}{1-kr^2} 
+r^2 d\Omega_{(2)}^2 \right) \,,
\ee
while the new Brans-Dicke field~(\ref{symmetry2}) is 
\be
\tilde{\phi}(\tau) = \left( \alpha q+1 \right)^{ 
\frac{q(1-2\alpha)}{\alpha q+1}} \phi_0^{ 1-2\alpha} \tau^{ 
\frac{q(1-2\alpha)}{\alpha q+1} } \,.
\ee
One can write 
\be
\tilde{S}( \tau) = S_0 \tau^{\tilde{p}} \,, \;\;\;\;\;\;\;\;\; 
\tilde{\phi}( \tau) = \tilde{\phi}_0 \tau^{\tilde{q}} \,,
\ee
where
\begin{eqnarray}
\tilde{p} & = & \frac{p+\alpha q}{\alpha q +1} \,, \label{tildep}\\
&&\nonumber\\
\tilde{q} & = & \frac{q(1-2\alpha)}{\alpha q +1} \,, \label{tildeq}\\
&&\nonumber\\
\tilde{\phi}_0 &=& \left( \alpha q+1 \right)^{ \frac{q(1-2\alpha)}{\alpha q 
+1}} \phi_0^{1-2\alpha} \,.
\end{eqnarray}
As a special situation, we discuss the O'Hanlon and Tupper family of 
spatially flat solutions of vacuum Brans-Dicke cosmology given, for $k=0$,  
by \cite{OHanlonTupper} 
\begin{eqnarray}
q_{\pm} &=& \frac{1}{3\omega+4} \left( \omega +1 \pm \sqrt{ 
\frac{2\omega+3}{3} } \right)\,,\\
&&\nonumber\\
p_{\pm} &=& \frac{ 1\mp \sqrt{3(2\omega+3)}}{3\omega+4} \,,
\end{eqnarray}
whose exponents satisfy the relation
\be
3q_{\pm}+p_{\pm}=1 \,.
\ee
Equations~(\ref{tildep}) and (\ref{tildeq}) give 
\be
3\tilde{q}+\tilde{p}  = \frac{1-5\alpha q}{\alpha q+1} \,,
\ee
which is, in general, different from unity, hence the new solution 
generated here is not of the O'Hanlon and Tupper form.

\subsection{Exponential solutions with linear potential} 

Instead of power-law solutions, we now use exponential solutions of vacuum 
Brans-Dicke theory with a linear potential $V(\phi)=\Lambda \phi$, which 
amounts to introducing a cosmological constant in this theory. The 
spatially flat family of solutions 
\begin{eqnarray}
S_{\pm}(t) &=& S_0 \exp\left[ \pm (\omega+1) \sqrt{ \frac{ 
2\Lambda}{(2\omega+3)(3\omega+4)}} \, t \right] \,,\nonumber\\
&&\\
\phi_{\pm}(t) &=& \phi_0 \exp\left[ \pm \sqrt{ \frac{ 
2\Lambda}{(2\omega+3)(3\omega+4)}} \, t \right] \,,
\end{eqnarray}
with $S_0, \phi_0$ constants, are well-known attractors in phase space 
\cite{81, 744, 544}. By performing the conformal 
transformation~(\ref{symmetry1}) 
one obtains
\begin{eqnarray}
d\tilde{s}^2 &=& \phi^{2\alpha} ds^2 =-\mbox{e}^{ \pm 2\alpha 
\sqrt{ \frac{2\Lambda}{(2\omega+3)(3\omega+4)}}\, t} dt^2 \nonumber\\
&&\nonumber\\
&\, & + S_0^2 \exp\left[ \pm 2 
(\omega+1+\alpha) \sqrt{ \frac{ 2\Lambda}{(2\omega+3)(3\omega+4)}} \, t 
\right] \nonumber\\
&&\nonumber\\
&\, & \cdot \left( dr^2+r^2 d\Omega_{(2)}^2 \right) \,.
\end{eqnarray}
The comoving time in the tilded world is 
\be
\tau =\frac{
\mbox{e}^{ \pm \alpha \sqrt{ \frac{ 2\Lambda}{(2\omega+3)(3\omega+4)} 
} \, t} }{ \pm \alpha \sqrt{\frac{ 2\Lambda}{(2\omega+3)(3\omega+4)}
} } + \mbox{const.}  \label{world}
\ee
One must make sure that $t$ and $\tau$ have the same direction. By 
choosing the positive sign this property follows trivially and $\tau=0$ 
corresponds to $t\rightarrow -\infty$. If the negative sign is taken, one 
can choose the integration constant so that 
\be
\tau= \frac{1}{\alpha} \sqrt{ \frac{(2\omega+3)(3\omega+4)}{2\Lambda}} 
\left( 1- \mbox{e}^{ -\alpha \sqrt{ 
\frac{2\Lambda}{(2\omega+3)(3\omega+4)}} \, t} \right) \,. \label{tauneg}
\ee

Consider first the solution with positive sign, which is 
rewritten as 
\be
d\tilde{s}^2 =-d\tau^2 +\tilde{S}^2(\tau)  \left( dr^2 +r^2 
d\Omega_{(2)}^2 \right) \,,\label{ancora}
\ee
where
\begin{eqnarray}
\tilde{S}(\tau)&=&\tilde{S}_0 \tau^{ \frac{\omega+1+\alpha}{\alpha} } 
\equiv \tilde{S}_0  \tau^{\tilde{p}} \,,  \label{stocz1}\\
&&\nonumber\\
\tilde{\phi}(\tau)&=&\phi^{ 1-2\alpha}= \tilde{\phi}_0  \tau^{ 
\frac{1-2\alpha}{\alpha} } \equiv \tilde{\phi}_0  \tau^{ 
\tilde{q}}\,,\label{stocz2}
\end{eqnarray}
where
\begin{eqnarray}
\tilde{S}_0 &=& S_0 \left[ \alpha 
\sqrt{\frac{2\Lambda}{(2\omega+3)(3\omega+4)} } \right]^{ 
\frac{\omega+1+\alpha}{\alpha}} \,,\\
&&\nonumber\\
\tilde{\phi}_0 &=& \phi_0^{ 1-2\alpha} \left( \alpha  
\sqrt{\frac{2\Lambda}{(2\omega+3)(3\omega+4)}} 
\right)^{\frac{1-2\alpha}{\alpha}} \,,
\end{eqnarray}
and 
\be
3\tilde{q}+\tilde{p} = \frac{ \omega +4-5\alpha}{\alpha} 
\ee
(which, in general, is not equal to~1). The scalar field potential is now, 
according to Eq.~(\ref{newV}), 
of the power-law form
\be
\tilde{V}(\tilde{\phi})= \tilde{\phi}^{\frac{-4\alpha}{1-2\alpha}} \Lambda 
\phi = \Lambda \tilde{\phi}^{ \frac{1-4\alpha}{1-2\alpha}} \,.\label{cz20}
\ee
Using Eq.~(\ref{newomega}), the scale factor is written in terms of the 
new Brans-Dicke coupling $ \tilde{\omega}$ as 
\be
\tilde{S}(\tau) = \tilde{S}_0 \, \tau^{ 
\frac{1-2\alpha}{\alpha} \left[  
\tilde{\omega}(1-2\alpha) +1-3 \alpha \right]} \,.
\ee
In this case, the symmetry~(\ref{symmetry1}), (\ref{symmetry2}) transforms 
an exponential solution into a power-law one corresponding to a different 
power-law potential.

By choosing the negative sign in Eq.~(\ref{world}), we have instead the 
line element~(\ref{stocz1}) with  
\be 
S(\tau) =S_0 
\left( 1- \alpha \sqrt{ \frac{2\Lambda}{(2\omega+3)(3\omega+4)}} \, \tau 
\right)^{\frac{\omega+1+\alpha}{\alpha} } 
\ee
 but $\tau$ is now given by Eq.~(\ref{tauneg}) and 
\begin{eqnarray}
\tilde{\phi}(\tau) &=& \phi_0^{ 1-2\alpha} \left( 1-\alpha  
\sqrt{\frac{2\Lambda}{(2\omega+3)(3\omega+4)}} \, \tau  
\right)^{\frac{1-2\alpha}{\alpha}} \,. \nonumber\\
&& 
\end{eqnarray}
The scalar field potential is again~(\ref{cz20}).

\section{A new family of spherical, time dependent solutions}
\label{sec:4}
\setcounter{equation}{0}

In this section we use the symmetry transformation to generate a new 
time-dependent solution of Brans-Dicke theory from a static one used as a 
seed. 

A spherically symmetric and time-dependent solution of Jordan frame 
vacuum Brans-Dicke theory, which is conformal to the Fonarev 
spacetime\footnote{The Fonarev solution of GR, in turn, is conformal 
to the 
Fisher-Buchdahl-Janis-Newman-Winicour-Wyman scalar field
solution of the Einstein equations \cite{Fisher}.} \cite{Fonarev, 
MaedaFonarev}, 
was found recently in \cite{confonarev}. The line element and Brans-Dicke 
field are  
\begin{eqnarray}
ds^2 &=& - A(r)^{ \frac{1}{\sqrt{1+4d^2}} 
(2d-\frac{1}{\sqrt{|2\omega+3|}})} 
\, \mbox{e}^{ 4dat( 2d-\frac{1}{ \sqrt{|2\omega+3|}}) } dt^2 \nonumber\\
&&\nonumber\\
&\, & +\, \mbox{e}^{ 2at ( 1-\frac{2d}{\sqrt{|2\omega+3|}} )} \left[ 
A(r)^{\frac{-1}{\sqrt{1+4d^2}}(2d+\frac{1}{\sqrt{|2\omega+3|}})} dr^2 
\right.\nonumber\\
&&\nonumber\\
&\, & \left. +
A(r)^{1- \frac{1}{\sqrt{1+4d^2}} (2d+\frac{1}{\sqrt{|2\omega+3|}})} r^2 
d\Omega_{(2)}^2 \right] \,,\label{confonarev1}
\end{eqnarray}
\be
\phi(t,r)=\phi_0 \, \mbox{e}^{ \frac{4dat}{\sqrt{|2\omega+3|}}} A(r)^{ 
\frac{1}{ \sqrt{|2\omega+3|(1+4d^2)}} } \,,\label{confonarev2}
\ee
where 
\be
A(r) =1-\frac{2m}{r} \,,
\ee
\be 
V(\phi)=V_0 \phi^{\beta} \,,  \;\;\;\;\;\;
\beta=2\left(1-d\sqrt{|2\omega+3|}\right) \,,
\ee
and where $m>0$, $a, d$ are parameters of the family of solutions, while 
$\omega\neq -3/2$ and $\phi_0>0$ is another 
constant related to initial conditions. 

We use a special case of this family as the seed to generate a new 
family of solutions of vacuum Brans-Dicke gravity. Assuming $a\neq 0$, the 
time dependence of the 
geometry~(\ref{confonarev1}) is 
eliminated if the parameter $d$ is simultaneously equal to $\left( 
2\sqrt{|2\omega+3|}\right)^{-1}$ and to $\sqrt{|2\omega+3|}/2$, which is 
achieved if $\omega=-1$ or if $\omega=-2$. In these cases, however, the 
scalar field~(\ref{confonarev2}) remains time-dependent, while 
$\beta=1$ and the scalar 
field potential reduces to the linear $V(\phi)=V_0\phi$, which is 
equivalent to introducing a cosmological constant in the Brans-Dicke 
action. Since $\phi>0$, this potential is effectively bounded from below.  
A scalar field which does not share the symmetries of the spacetime metric 
is currently of considerable interest because it is used in Brans-Dicke, 
Galileon, and Horndeski gravity as an 
ingredient to circumvent 
(\cite{circumvent}, see also \cite{also})  well-known no-hair theorems for 
black holes \cite{Hawking, 
SotiriouFaraoni, BhattaRomano}.  

Our starting point is 
\begin{eqnarray}
ds^2 &=& -dt^2 +A(r)^{-\sqrt{2}} dr^2 +A(r)^{1-\sqrt{2}} r^2 
d\Omega_{(2)}^2 \,, \nonumber\\
&&\label{seed1}\\
\phi(t,r) &=&  \phi_0 \, \mbox{e}^{2at} A(r)^{1/\sqrt{2}} \,. 
\label{seed2}
\end{eqnarray}
This geometry is recognized as a special case of the Campanelli-Lousto 
geometry of Brans-Dicke theory. The general Campanelli-Lousto solution has 
the form \cite{CampanelliLousto} 
\begin{eqnarray}
ds^2_{CL} &=& -A(r)^{b_0+1}dt^2 + A(r)^{-a_0-1} dr^2  \nonumber\\
&&\nonumber\\
&\, & +A(r)^{-a_0}r^2 d\Omega_{(2)}^2 \,, \label{CL1}\\
&&\nonumber\\
\phi_{CL}(r) &=& \phi_0 A(r)^{\frac{a_0-b_0}{2}} \,,\label{CL2}
\end{eqnarray}
where $a_0$ and $b_0$ are two parameters, only one of which is 
independent, and are related to the Brans-Dicke 
coupling by 
\be
\omega(a_0, b_0) =\frac{-2\left(a_0^2 +b_0^2 -a_0b_0 
+a_0+b_0\right)}{\left(a_0-b_0 \right)^2} \,. \label{omegaab}
\ee
The line element~(\ref{seed1}) and scalar field~(\ref{seed2}) are  
reproduced if $\left(a_0, b_0 
\right)=\left( \sqrt{2} -1, -1 \right)$, while Eq.~(\ref{omegaab}) gives 
back $\omega =-1$ (but not the value $\omega=-2$ because  the 
Campanelli-Lousto solution holds for $\omega>-3/2$). Brans-Dicke gravity 
with this value of the 
Brans-Dicke coupling corresponds to the low-energy limit of bosonic string 
theory \cite{bosonic}, so it is plausible that the 
spacetime~(\ref{seed1}), (\ref{seed2}) has some stringy analogue.  
Although it was originally presented as describing  
a black hole spacetime, it was attributed a zero temperature, and there 
are studies of the thermodynamics of such ``cold black holes'' 
\cite{coldBH}, the 
Campanelli-Lousto solutions can only describe wormholes or naked 
singularities but not black holes \cite{Vanzo}. 
When the parameter $a_0$ is positive, and therefore 
for the value $a_0=\sqrt{2}-1$ corresponding to~(\ref{seed1}), 
(\ref{seed2}), the 
Campanelli-Lousto geometry describes a wormhole with the throat located 
at the apparent horizon radius \cite{Vanzo}
\be 
r_H=(2+a_0)m =(\sqrt{2}+1)m\,,
\ee
which corresponds to the value 
\begin{eqnarray}
R_H &=& \left( 2+a_0 \right)^{\frac{a_0+2}{2}} a_0^{-a_0/2} m \nonumber\\
&&\nonumber\\
&=&  
\left( \sqrt{2}+1 \right)^{\frac{\sqrt{2}+1}{2}}  
(\sqrt{2}-1)^{\frac{1-\sqrt{2}}{2} }m  
\end{eqnarray}
of the areal 
radius $ R(r)= rA(r)^{-a_0/2} $. By applying the symmetry 
transformation~(\ref{symmetry1}) and 
(\ref{symmetry2}), we obtain the new solution 
of vacuum Brans-Dicke theory with scalar field potential
\be
\tilde{V}(\tilde{\phi}) = V_0 \,  
\tilde{\phi}^{\frac{1-4\alpha}{1-2\alpha}} 
\ee
and Brans-Dicke coupling
\be
\tilde{\omega} = \frac{6\alpha(1-\alpha)-1}{ (1-2\alpha)^2} 
\ee
given by
\begin{eqnarray}
d\tilde{s}^2 &=&  -\mbox{e}^{4\alpha at} A(r)^{\alpha \sqrt{2}} dt^2 
+\mbox{e}^{4\alpha at} \left[ A(r)^{ -\sqrt{2} (1-\alpha)} dr^2 
\right.\nonumber\\
&&\nonumber\\
&\, & \left. + 
A(r)^{1-\sqrt{2}(1-\alpha)}r^2 d\Omega_{(2)}^2 \right] \,,\label{temp1}\\
&&\nonumber\\
\tilde{\phi}(t,r) &=& \tilde{\phi}_0 \, \mbox{e}^{2a(1-2\alpha)t} 
A(r)^{\frac{1-2\alpha}{\sqrt{2}} } \,, \;\;\;\;\;\;
\tilde{\phi}_0 = \phi_0^{1-2\alpha} \,.\label{temp2}
\end{eqnarray}
(Incidentally, applying the symmetry transformation to the general 
Campanelli-Lousto spacetime~(\ref{CL1}), (\ref{CL2}) does not produce 
another Campanelli-Lousto solution.)  Equations~(\ref{temp1}) and 
(\ref{temp2}) describe a 3-parameter family of 
solutions parametrized by $\left(m, a, \alpha \right)$.
If $a=0$ the time dependence disappears and this solution reduces again to 
a Campanelli-Lousto geometry with new parameters $\left(a_0, b_0 
\right)=\left( 
\sqrt{2}(1-\alpha)-1 , \alpha \sqrt{2}-1\right)$. Equation~(\ref{omegaab}) 
then gives $\tilde{\omega} = (-6\alpha^2+6\alpha-1)(1-2\alpha)^{-2}$ 
which, of course, matches Eq.~(\ref{newomega}) for $\omega=-1$. 

If $a\neq 0$, the new time coordinate 
\be
\tau = \frac{\mbox{e}^{2\alpha at}}{ 2\alpha a} 
\ee
transforms the spacetime $\left( \tilde{g}_{ab}, \tilde{\phi} \right)$  
into
\begin{eqnarray}
d\tilde{s}^2 &=&  - A(r)^{\alpha \sqrt{2}} d\tau^2 
+\left( 2\alpha a\tau\right)^2  \left[ A(r)^{ -\sqrt{2} (1-\alpha)} dr^2 
\right.\nonumber\\
&&\nonumber\\
&\, & \left. + 
A(r)^{1-\sqrt{2}(1-\alpha)} r^2 d\Omega_{(2)}^2 \right] \,,\label{cz1}\\
&&\nonumber\\
\tilde{\phi}(\tau,r) &=&\phi_* \, \tau ^{ \frac{1-2\alpha}{\alpha} } 
A(r)^{\frac{1-2\alpha}{\sqrt{2}} } \,, \;\;\;\;\;\;
\tilde{\phi}_*  =  \left[ \left( 2\alpha a\right)^{1/\alpha}  
\phi_0\right]^{1-2\alpha} 
\,.\nonumber\\
&& \label{cz2}
\end{eqnarray}
By taking the limit $m\rightarrow 0$, this geometry reduces to the 
spatially flat FRLW universe
\be
d\tilde{s}_{(1)}^2 = - d\tau^2 
+\left( 2\alpha a\tau\right)^2  \left( dr^2 + r^2 d\Omega_{(2)}^2 \right) 
\ee
with comoving time $\tau$, linear scale factor $S(\tau)=2\alpha a \tau$, 
and scalar field
\be
\tilde{\phi}_{(1)} (\tau,r) =\phi_* \, \tau ^{ \frac{1-2\alpha}{\alpha} } 
\ee
(this is not a O'Hanlon and Tupper universe). The same line element and 
Brans-Dicke scalar are obtained asymptotically 
for $r\gg 2m$. Therefore, the new solution is interpreted as a spherical 
inhomogeneity in a spatially flat FLRW universe, with the scalar $\phi$ 
behaving asymptotically as a perfect fluid with equation of state 
$P=-\rho/3$. It is not trivial to establish the nature of the central 
inhomogeneity. Since the Campanelli-Lousto geometry with parameter 
$a_0>0$, to which the new 
solution is conformal, can describe only a  wormhole \cite{Vanzo}, one 
would naively expect that its conformal 
cousin describes the same type of solutions. While this is indeed the 
case for spherical geometries resulting from the conformal 
transformation of a wormhole or a naked singularity with static conformal 
factor \cite{exoticbh}, a time-dependent conformal factor may change this 
picture. As an example, a time-dependent conformal transformation 
of the static 
Fisher solution (which contains a naked singularity \cite{Fisher}), 
produces the Husain-Martinez-Nu\~nez solution of the Einstein equations in 
which a central singularity is covered by a black hole apparent horizon 
for part of the history of this spacetime (according to the comoving time 
of the FLRW background)  \cite{HMN}.  A detailed study of the physical 
interpretation of the solution~(\ref{cz1}) and (\ref{cz2}) will be 
reported elsewhere.

\section{Generating new axially symmetric solutions} 
\label{sec:5} 
\setcounter{equation}{0}

The solution-generating technique can be applied to cylindrically 
symmetric spacetimes. Since the Brans Dicke action remains invariant when 
an electromagnetic field is added to it,  we present  a new solution 
generated by using a cylindrically symmetric electrovacuum Brans Dicke 
spacetime as a seed \cite{Ahmet}. The latter contains only an azimuthal 
magnetic  field $B$ and  the line element takes the form 
\begin{eqnarray}
ds^2=&&(1+c^2r^p)^2\left[ r^{2(q-d)}(-dt^2+dr^2) \right. \nonumber\\
&&\nonumber\\
&& \left. + W_0^2 r^{2(k-d)} d\theta^2 \right]  
+\frac{r^{2d}}{(1+c^2r^p)^2}dz^2  \label{clylindrical1}
\end{eqnarray}
in cylindrical coordinates $\left( t, r, \theta, z \right)$, where 
\begin{eqnarray}
&&p=2d-k+1 \,, \\
&&\nonumber\\
&& q(\omega) = d(d-k+1)+\frac{\omega}{2} (k-1)^2+k(k-1) \,\label{qw}.
\end{eqnarray} 
The scalar and the magnetic field are
\begin{eqnarray}
&&\phi(r) = \phi_0\, r^{1-k} \,, \label{clphi}\\
&&\nonumber\\
&& B_r=B_z=0 \,, \;\;\;\;\; B_\theta (r)=\pm\frac{\sqrt{\phi_0} \, 
c\,p\,r^{p-1}}{(1+c^2r^p)^2} 
\,,\label{clB}
\end{eqnarray}
respectively. $\phi_0$ is a positive constant, while  
the constant $c$ is related to the coupling of the electromagnetic field 
to the current \cite{Ahmet}. If $c=0$, (\ref{clylindrical1})-(\ref{clB}) 
becomes the electrovacuum solution of \cite{Arazi} with the Levi-Civita 
geometry. When $c=0$, the constant $W_0$ introduces a conical 
singularity (if $W_0^2 \neq 1$), while the constant $d$ is related with 
the energy density of the electromagnetic source. For $d=0$, the line 
element~(\ref{clylindrical1}) describes a  
cosmic string in Brans-Dicke-Maxwell theory in which the $z$-axis carries 
a  current \cite{Vilenkin}. $k$ is a free parameter and, for $k=1$, this  
spacetime  reduces to an Einstein-Maxwell solution \cite{KSMcCH, 
Bonnor}. If $c=0$ and $k=1$ simultaneously, this geometry reduces to the 
Levi-Civita solution of the vacuum Einstein equations \cite{LeviCivita}. 
Equations~(\ref{clylindrical1})-(\ref{clB}) describe a family of 
solutions  
characterized by the four parameters $\left( \omega, k,d,c \right)$, where 
$\omega$ is a parameter of the theory.

After performing the conformal transformation~(\ref{symmetry1}), the line 
element~(\ref{clylindrical1}) is 
\begin{eqnarray}
d\tilde{s}^2 &=& (1+c^2r^p)^2\bigg{\{} r^{2 \left[ q-d   
+ \alpha(1-k) \right] } \left( -dt^2+dr^2 \right) 
\nonumber\\
&&\nonumber\\
&\, &  + W_0^2 r^{ 2[k-d+\alpha(1-k)]} d\theta^2\bigg{\}} +\frac{r^{2[ d+\alpha(1-k)]}}{(1+c^2r^p)^2}dz^2\, ,\nonumber\\ 
\label{clylindrical2}
\end{eqnarray}
 where $q$ is simply given by (\ref{qw}). In the conformally 
transformed  frame, one can also express this equation in terms of new 
Brans-Dicke 
parameter $\tilde{\omega}$ and of $\tilde{q} \equiv 
q(\tilde{\omega})$. Then Eq.~(\ref{clylindrical2})  becomes
\begin{eqnarray}
d\tilde{s}^2 &=& (1+c^2r^p)^2\bigg{\{} r^{2 \left[ \tilde{q} 
	+\alpha(\alpha-1)(k-1)^2(2\tilde{\omega}+3)   -d   
	+ \alpha(1-k) \right] } 
\nonumber\\
&&\nonumber\\
&\, &  \cdot \left( -dt^2+dr^2 \right) 
+ W_0^2 r^{ 2[k-d+\alpha(1-k)]} d\theta^2\bigg{\}} \nonumber\\
&&\nonumber\\ 
&\, & +\frac{r^{2[ d+\alpha(1-k)]}}{(1+c^2r^p)^2}dz^2 \,, 
\label{clylindrical3}
\end{eqnarray}
 and the new scalar field reads
\begin{equation}
\tilde{\phi}=\tilde{\phi}_0\, r^{(1-2\alpha)(1-k)}
\quad \text{with} \quad \tilde{\phi}_0=\phi_0^{1-2\alpha} \,.
\end{equation}
We have  a new family of solutions characterized by the five parameters 
$\left( \tilde{\omega}, \alpha, k,c,d \right)$ (where $\tilde{\omega}$ is 
a parameter of the theory).

A conical singularity is present in both geometries (\ref{clylindrical1}) 
and (\ref{clylindrical2}), but we can eliminate it in the case $c=0$. 
Choosing  $k-d=1$ and $W_0^2=1$ in Eq.~(\ref{clylindrical1}) reduces  the 
line element to
\begin{eqnarray}
ds^2 =&&(1+c^2r^{p})^2\bigg{\{} r^{p^2 (\omega +2) }     
\left( -dt^2+dr^2 \right) +r^2 d\theta^2\bigg{\}}\nonumber\\
&&\nonumber\\
&&+\frac{r^{2p}}{(1+c^2r^p{})^2}dz^2 \,, \label{clylindrical3}
\end{eqnarray}
while the scalar field is $\phi(r)=\phi_0 r^{-p}$ and $p=d=k-1$. In 
the limit $c\rightarrow 0$, the conical singularity disappears.

For the new solution~(\ref{clylindrical2}), we suggest that $W_0^2=1$ 
and  $k-d+\alpha(1-k)=1$, then the metric becomes
\begin{eqnarray}
d\tilde{s}^2=&& \left( 1+c^2r^{p_0} \right)^2\left[  
r^{ p_0^2 (\tilde{\omega}+2) }  \left( -dt^2+dr^2 \right) +r^2 
d\theta^2\right] \nonumber\\
&&\nonumber\\
&&+\frac{r^{2p_0} }{(1+c^2r^{p_0})^2} dz^2 \,,    \label{clylindrical4}
\end{eqnarray}
where now $\tilde{\phi}=\tilde{\phi_0}r^{-p_0}$ and 
$p_0 =(k-1)(1-2\alpha)$. Of course, Eq.~(\ref{clylindrical3}) is obtained 
as the $\alpha\rightarrow 0$ limit of Eq.~(\ref{clylindrical4}).

Let us discuss now the limit to GR. For both the seed spacetime and the 
new spacetime, the limit $k\rightarrow 1$ 
(which implies $p_0 \rightarrow 0$) reproduces GR and, if we choose 
$W_0^2 \neq 1$ and we rescale the coordinates according to 
\begin{eqnarray}
t &\rightarrow & \bar{t} = \left( 1+c^2 \right) t \,, \\
&&\nonumber\\
r &\rightarrow & \bar{r}= \left( 1+c^2 \right) r \,, \\
&&\nonumber\\
z &\rightarrow &  \bar{z}= \frac{z}{1+c^2}  \,, 
\end{eqnarray}
we obtain the 
cosmic string geometry  
\begin{equation}
ds^2=-d\bar{t}^2+d\bar{r}^2+d\bar{z}^2+W_0^2 \bar{r}^2 d\theta^2 
\label{string} 
\end{equation}
with zero magnetic field. Rescaling the coordinates is {\em de 
facto} equivalent to setting $c=0$, but it is not necessary to do this 
explicitly: the magnetic field is automatically killed by making the 
scalar field constant in the GR limit of the new  solution (but not in 
the seed solution). The string~(\ref{string}) has 
linear energy density $ \mu $ along the $\bar{z}$-axis, where 
$W_0^2=1-4\mu$. If $W_0^2<1$, there is a 
deficit angle and $\mu>0$: the length of a circumference of  
radius $\bar{r}$ circling the 
$\bar{z}$-axis is $2\pi \bar{r} \sqrt{1-4\mu}$. If instead $ W_0^2>1$, 
there is an  excess angle corresponding to $\mu<0$ \cite{Matt, Volovik}.

\section{$f(R)$ gravity}
\label{sec:5bis}
\setcounter{equation}{0}

$f(R)$ theories of gravity are described by the action
\be
S_{f(R)}=\int d^4x \sqrt{-g} f(R) +S^{(matter)} \,,\label{f(R)action}
\ee
where $f(R)$ is a nonlinear function of the Ricci scalar and 
$S^{(matter)}$ is the matter action. The action describing the 
gravitational  
sector is equivalent to that of a subclass of scalar-tensor gravity 
\cite{reviews}. In 
fact, it can be shown that the gravitational 
action in~(\ref{f(R)action}) is equivalent to that of  a Brans-Dicke 
theory 
with Brans-Dicke field $\phi =f'(R)$, Brans-Dicke coupling $\omega=0$, and 
scalar field 
potential
\be
V(\phi) =\phi R -f(R) \bigg|_{R=R(\phi)} \,, \label{f0}
\ee
where $R$ is to be understood as a function of the scalar degree of 
freedom  $\phi=f'(R)$. It is then natural to ask whether the 
1-parameter symmetry group of transformations (\ref{symmetry1}), 
(\ref{symmetry2}) of Brans-Dicke theory generates symmetries of $f(R)$ 
gravity, $f(R) \rightarrow \tilde{f}( \tilde{R})$. Answering this question 
turns out to be complicated. First of all, in order to keep the 
equivalence between $\tilde{f}( \tilde{R})$ gravity and Brans-Dicke 
theory, it must be
\begin{eqnarray}
\tilde{\phi} &=& \frac{ d\tilde{f}}{d\tilde{R}} \,, \label{f1}\\
&&\nonumber\\
\tilde{\omega} &=& 0 \,,\label{f2}\\
&&\nonumber\\
\tilde{V}( \tilde{\phi} ) &=& \tilde{\phi} \tilde{R} -\tilde{f}( 
\tilde{R}) \,. \label{f3}
\end{eqnarray}
 Equation~(\ref{f2}) fixes the parameter $\alpha$ of the transformation to 
be  $\alpha=1$ and, at best, a single symmetry transformation of the 
$f(R)$ theory exists and not an entire 1-parameter group. 
Assuming 
$\alpha=1$, Eq.~(\ref{f1}) gives $\tilde{\phi}=\phi^{-1}$. It is more 
complicated to enforce Eq.~(\ref{f3}). Using Eq.~(\ref{f1}), the 
latter becomes
\be
\tilde{V}( \tilde{\phi}) = \tilde{\phi} \tilde{R} -\tilde{f}(\tilde{R}) 
\,;
\ee
it must be
\be
\tilde{\phi}^3 \left( \tilde{R}-\tilde{\phi}f(R) \right)=\tilde{\phi} 
\tilde{R}-\tilde{f}( \tilde{R}) \label{f6} 
\ee
if the function $\tilde{f}(\tilde{R})$ is going to be generated by the 
operation~(\ref{symmetry1}), (\ref{symmetry2}) with $\alpha=1$. By 
remembering the transformation property of the Ricci scalar under the 
conformal transformation $g_{ab} \rightarrow \tilde{g}_{ab} =\phi^2 
g_{ab}$ corresponding to $\alpha=1$ 
\be
\tilde{R}=\frac{1}{\phi^2} \left( R-\frac{6\Box\phi}{\phi} \right) 
\,,\label{f7}
\ee
\cite{Synge, Waldbook, Carroll, mybook} and using the field 
equation~(\ref{BDfe2}) {\em in vacuo} or 
electrovacuo 
($T=0$), one obtains
\be
\tilde{R}= \frac{6}{\left( 2\omega+3 \right)\left[ f'(R) \right]^2} \left[ 
\frac{2 f(R)}{f'(R)} -R \right] \,.\label{f8}
\ee
If this equation could be inverted to express $R=R( \tilde{R})$ 
explicitly, the result could then be substituted into Eq.~(\ref{f7}),  
producing
a nonlinear ordinary differential equation for the function $\tilde{f}( 
\tilde{R})$ satisfied by all $\tilde{f}( \tilde{R})$ theories generated 
by the Brans-Dicke symmetry. Moreover, one could then write down   
explicitly the form of the function $\tilde{f}( \tilde{R})$. In practice, 
these steps cannot be performed. The root of the problem lies in the fact 
that the potential~(\ref{f0}) of the scalar $\phi$ of $f(R)$ gravity is 
not explicit (because one cannot invert explicitly the relation 
$\phi=f'(R)$ in order to obtain $R=R(\phi)$). Thus, in general, the 
question of whether the Brans-Dicke symmetry (\ref{symmetry1}), 
(\ref{symmetry2}) generates  a symmetry of $f(R)$ gravity cannot be 
answered. However, we can propose a special solution to this problem.

A solution is found for the special choice $f(R)=R^n$ of the function 
$f(R)$, which has been the subject of an extensive literature 
\cite{SolarSystem, powerlaw}.  In this case we have 
\be
\phi = n R^{n-1} \,, \;\;\;\; V(\phi)=(n-1)R^n \,, \;\;\; 
\tilde{\phi}= \frac{ R^{1-n}} {n}  \,,
\ee
and Eq.~(\ref{f6}) becomes
\be
\tilde{R} \, \frac{ d\tilde{f} }{ d\tilde{R}} - \tilde{f}(\tilde{R}) 
= \frac{n-1}{n^4} \, R^{4-3n} \,.
\ee
In vacuo or electrovacuo we have
\be
\frac{\Box \phi}{\phi} = \frac{1}{2\omega+3} \left( \frac{2f}{\phi} - R 
\right) 
\ee
and
\be
\tilde{R}=\frac{\left( 2\omega n+9n -12\right) }{n^3 \left( 2\omega+3 
\right)} \, R^{3-2n} \,,
\ee
which leads to
\be
R =\left[ \frac{n^3 \left(2\omega+3 \right)}{2\omega n+9n -12} \, \tilde{R} 
\right]^{\frac{1}{3-2n}}   \,. \label{f9} 
\ee
Equation~(\ref{f6}) can then be written as the first order ordinary 
differential equation for $\tilde{f}(\tilde{R})$ 
\be
\tilde{R} \, \frac{ d\tilde{f}}{d\tilde{R}} - \tilde{f}( \tilde{R}) - \mu 
\tilde{R}^{ \frac{3n-4}{2n-3}} =0 \,,
\ee
where
\be 
\mu=
\left[ \frac{n^3 \left(2\omega+3 \right)}{2\omega n+9n -12} 
\right]^{ \frac{3n-4}{2n-3}} \left( \frac{n-1}{n^4} \right) \,.
\ee
A solution is
\begin{eqnarray}
\tilde{f}( \tilde{R}) &=& \alpha \tilde{R}^{ \tilde{n}} \,,\\
&&\nonumber\\
\tilde{n} &=& \frac{3n-4}{2n-3}   \,, \label{f10}\\
&&\nonumber\\
\alpha &=& \left( \frac{2n-3}{n-1} \right) \mu \,.
\end{eqnarray}
General theoretical constraints on any $f(R)$ gravity theory 
\cite{reviews} are $f'>0$, 
which guarantees that the graviton carries positive kinetic energy, and 
$f''\geq 0$, which guarantees local stability \cite{mattmodgrav}. For  
$f(R)=R^n$ , these constraints imply $n\geq 1$, and here we discard the 
value 
$n=1$ corresponding to GR, in which case $\phi=$~const. and the 
symmetry~(\ref{symmetry1}), (\ref{symmetry2}) degenerates. Applying 
these constraints to the theory described by $\tilde{f}(\tilde{R})$ (but 
not necessarily to the ``seed'' theory $f(R)=R^n$) implies $\tilde{n}>1$, 
which is compatible with the 
experimental constraint 
\be
\tilde{n}-1 =\left( -1.1\pm 1.2 \right) \cdot 10^{-5}
\ee
coming from Solar System experiments \cite{SolarSystem}.

For the special value $n=2$ of the exponent, and with vanishing 
Brans-Dicke parameter $\omega$,  
Eq.~(\ref{f10}) implies that $\tilde{n}=2$ as well, while $\mu=\alpha=1$ 
and $ \tilde{f}( \tilde{R} ) = \tilde{R}^2$. Therefore, the theory 
$f(R)=R^2$ is invariant under the transformation considered. 
Apart from the trivial case $n=1$ corresponding to GR, this is the only 
value of $n$ for which $\tilde{n}=n$.  While this 
particular $f(R)$ model is inconsistent with weak 
gravity experiments, it constitutes a good approximation of 
the Starobinsky inflationary model of the  
early universe $f(R)=R+ \alpha  R^2$  \cite{Starobinsky} in 
strong curvature regimes.

Other possible solutions of 
Eq.~(\ref{f6}) will be searched for in future work.

\section{Conclusions}
\label{sec:6}
\setcounter{equation}{0}

The symmetry group of a physical theory discloses some of its fundamental 
features. We have reported a symmetry of Brans-Dicke theory {\em in 
vacuo}, 
electrovacuo, and in the presence of conformally invariant matter, which 
includes a radiation fluid important in the radiation era of cosmology 
and in star models.  This symmetry consists of a restricted conformal 
invariance of 
the theory under very specific conformal transformations accompanied by 
nonlinear redefinitions of 
the Brans-Dicke scalar, and it is not to be confused with the usual 
conformal transformation from the Jordan to the Einstein frame of 
scalar-tensor gravity and string theories.  The symmetry was reported long 
ago \cite{myBDlimit} and was used to investigate anomalies \cite{failure} 
in the limit of Brans-Dicke gravity  to GR (\cite{myBDlimit}, see 
\cite{newrefs} for further developments). Here the symmetry of 
\cite{myBDlimit} is generalized  to include the 
case in which the Brans-Dicke scalar $\phi$ is endowed with an 
arbitrary potential $V(\phi)$ and conformally invariant matter is 
possibly present.

We propose a novel use of this symmetry as a solution-generating 
technique, starting from a known solution of the theory used as a seed. As 
examples, we have reported new solutions of FLRW Brans-Dicke cosmology in 
the presence of a cosmological constant, a new 3-parameter family of 
spherical, time-dependent vacuum solutions (which are rare in the 
literature, contrary to static spherical solutions which are much easier 
to find), and a new family of cylindrically symmetric static electrovacuum 
solutions. The new spherical family, which is achieved using an 
$\omega=-1$ solution as a seed and probably has stringy analogues, looks 
rather intriguing and its physical interpretation will be studied in more 
detail in the future. The new symmetry offers some scope for extending 
studies of point-like Lagrangians with cyclic variables in FLRW cosmology 
and in the realm of static spherical solutions, which have been 
investigated extensively in the literature in relation with Noether 
symmetries in scalar-tensor gravities (see \cite{booksalv} for a summary).

As pointed out long ago by Dicke \cite{Dicke}, two conformal frames 
related by a conformal transformation are physically 
equivalent if the fundamental units of length, time, and mass  
(and all derived units) scale with 
appropriate powers of the conformal factor $\Omega$ of the 
transformation. In practice, it is not trivial to implement this 
requirement \cite{DeruelleSasaki, Vanzo}. This consideration is usually 
debated for the conformal transformation going from the Jordan to the 
Einstein 
conformal frame, which is excluded 
by our symmetry operation~(\ref{symmetry1}), (\ref{symmetry2}), but 
Dicke's argument 
is more general and, therefore, a conformal transformation accompanied by 
the appropriate 
rescaling of units would not generate physically new solutions according 
to Dicke's argument. However, in this work we have not implemented Dicke's 
rescaling of units but we have used instead the 
transformation~(\ref{symmetry1}), (\ref{symmetry2}) 
as a mathematical map. Therefore, the solutions obtained are 
indeed new solutions of the theory.

\begin{acknowledgments}

We are grateful to Nathalie Deruelle for a discussion  and to a 
referee for useful comments. D.K.C. thanks the Scientific and 
Technological Research Council of Turkey (T\"{U}B\.{I}TAK) for a 
postdoctoral fellowship through the Programme B\.{I}DEB-2219 and Nam{\i}k 
Kemal University for support. V.F. is supported by the Natural Sciences 
and Engineering Research Council of Canada (grant number 2016-03803), and 
all authors thank Bishop's University.

\end{acknowledgments}


\end{document}